\documentclass{article}

\usepackage{sao2}
\usepackage{graphicx}

\begin{document}

\title{On the GRB progenitors:
possible consequences for supernovae connection with
$\gamma$-ray bursts
}

\author{V.V. Sokolov}

\institute{\saoname}

\maketitle

\begin{abstract}
Unique data on $BVRI$ light curves of the optical transient (OT) of GRB 970508
obtained with the 6-m telescope have been interpreted in the framework
of the idea of a straightforward link between supernovae (SNe)
and long duration gamma-ray bursts (GRBs).
The effect must be maximum in the $I_{c}$ band as for OT GRB 970228.
The peak absolute ($M_{B}$) magnitude of the suggested SN
must be around -19.5 for the OT of GRB 970508.
So, in addition to the characteristic ``shoulders"
on the light curves of the OTs of GRB 970228, GRB 980326, 990712, 991208,
more evidence of the link between GRBs and Type Ib/c SNe (or core-collapse SNe)
was found, which could be an argument in favor of the idea of massive stars
as progenitors of long duration GRBs.
If all or the main part of long duration GRBs are associated with the SNe,
GRB host galaxies (for ground-based observations, at least)
must be dimmer than the peak magnitude of a SN.
If some GRB/SN relation really exists,
and if all or at least the main part
of long duration GRBs are associated with SNe,
then as a consequence we have a very strong $\gamma$-ray beaming
with a solid angle of up to $\Omega_{beam} \sim (10^{-5} - 10^{-6})\cdot 4\pi$.
Besides, the observations of K$_{\alpha}$ lines of iron
in the X-ray afterglow spectra of GRBs (970508, 970828, 991216, 000214)
and the observation of redshifted absorption feature of neutral iron (7.1 keV)
simultaneously with GRB 990705
are also evidence in favor of massive stars -- progenitors of GRBs.
\end{abstract}

\section{Introduction}\label{Intro}

The nature of $\gamma$-ray bursts is still a debated issue,
but the consensus is growing on the association of GRBs
with massive stars progenitors.
The long duraion GRBs are probably connected with the death of massive stars
and hence with a supernova-like explosion
(Paczin\'ski, 1999; MacFadyen \& Woosley, 1999, and references therein).
Over the past three years
this "hypernova" or "collapsar" model has gained strength,
especially because long duration GRBs have been found to occur
in star-forming regions of their galaxies --- the only places
where very massive stars live and die
(Fruchter et al., 1999; Hogg \& Fruchter, 1999; Bloom et al., 2000;
Sokolov et al., 2001).
The leading scenario now for long duration GRBs
is a sudden collapse of a massive rapidly rotating star.
May be merging neutron stars still remain a possibility
for the short GRBs, but this model (Lattimer \& Schramm, 1976)
has lost favor as an explanation for long duration GRBs,
those lasting more than 1 second.

This paper shows possible consequences for the scenario
explaining GRBs if massive stars are GRB progenitors actually.
In particular, as a result of evolution the massive stars
at a Wolf-Rayet phase (which is characterized by a strong stellar wind)
can produce rather dense envelopes around them
-- an effect of the WR stellar wind interacting with the interstellar
medium (Ramirez-Ruiz et al., 2001).

First we consider some features of $BVRI $ light curves
of such a well-studied object as the OT of GRB 970508
(Sokolov et al., 1998; Galama et al., 1998; Sokolov et al., 1999).
We begin the analysis of these features of the $BVRI$ light curves
when the GRB host galaxy is observed.
There is independent evidence in favor of the massive GRB progenitors:
at least in some GRBs a SN may be the origin of a red component
(or a shoulder) in the $BVRI$ light curves of a GRB afterglow,
detected several weeks after a GRB.
A possible contribution of a SN in the $BVRI$ light curves of GRB 970508
is considered.
Then, on the basis of the light curve of the OT of GRB 970508 and other
GRB transients with due regard
for the observable magnitudes of GRB host galaxies,
the possibility of the strong GRB beaming will be grounded.

The presence of a considerable component of the X-ray emission
in the radiation of BeppoSAX's GRBs
which is observed simultaneously with the long duration GRBs
(or a prompt GRB X-ray emission)
is very essential.
The Italian-Dutch X-ray mission BeppoSAX can provide not only arc minute
localization of GRBs, but also measurements of their spectra in a broad
(2 to 700 keV) energy band (Piro et al., 1998; Frontera et al., 2000).
The iron line features
(also detected in the X-ray afterglow with the ASCA and Chandra satellites)
would strongly suggest a massive stellar GRB progenitor too,
since these features can be direct evidence
for the existence of these dense envelopes in the vicinity of a pre-GRB object.
The account for a huge force of pressure on the environment
around a GRB-progenitor produced by a huge and highly collimated flux
of the prompt $\gamma$-ray and X-ray radiation
can considerably change our understanding of a GRB afterglow.

\section{The light curves: multiband photometry of the GRB970508 optical
counterpart}\label{Light_curves}

\begin{figure}
\vbox{
\includegraphics[width=\hsize,clip]{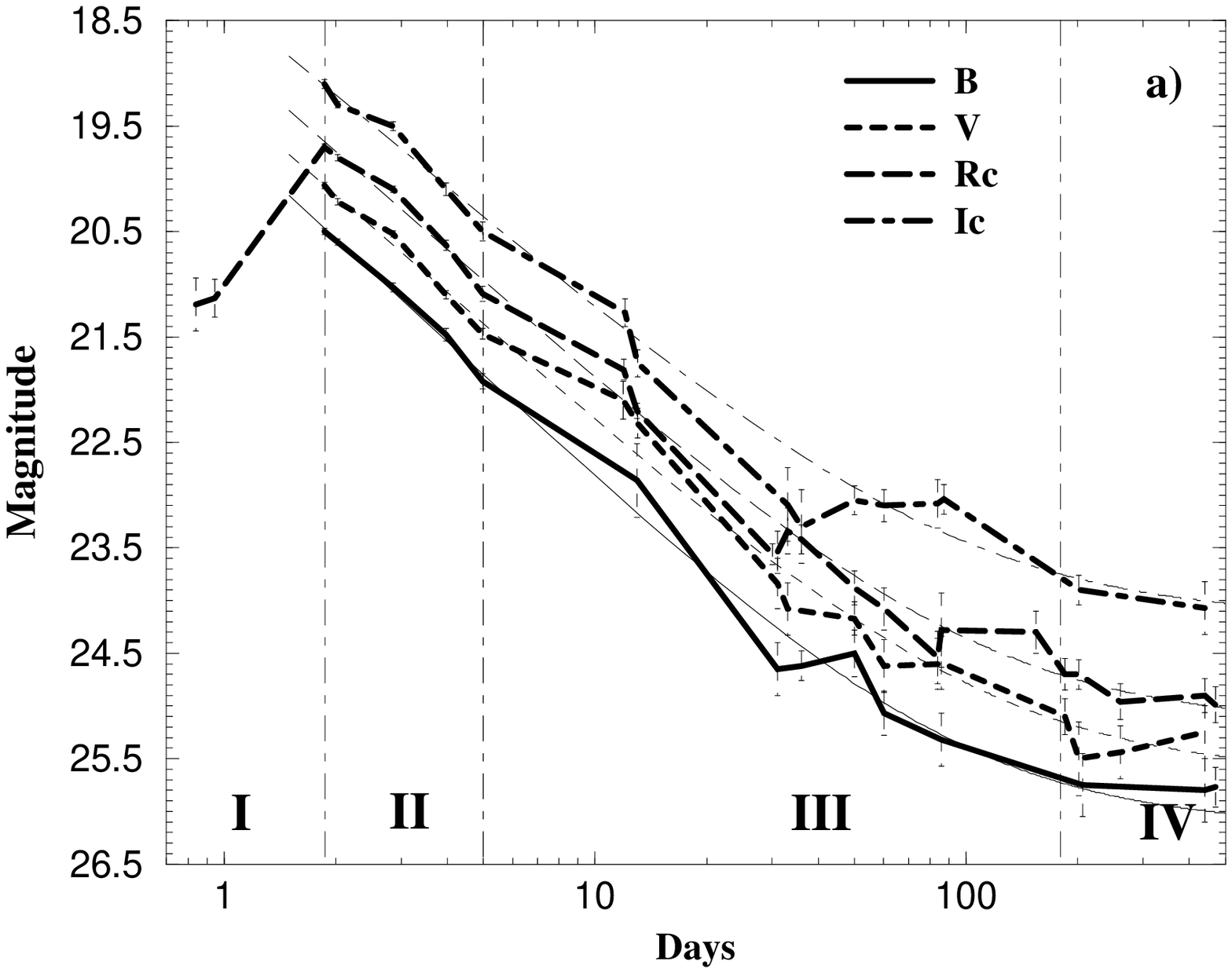}
\\
\includegraphics[width=\hsize,clip]{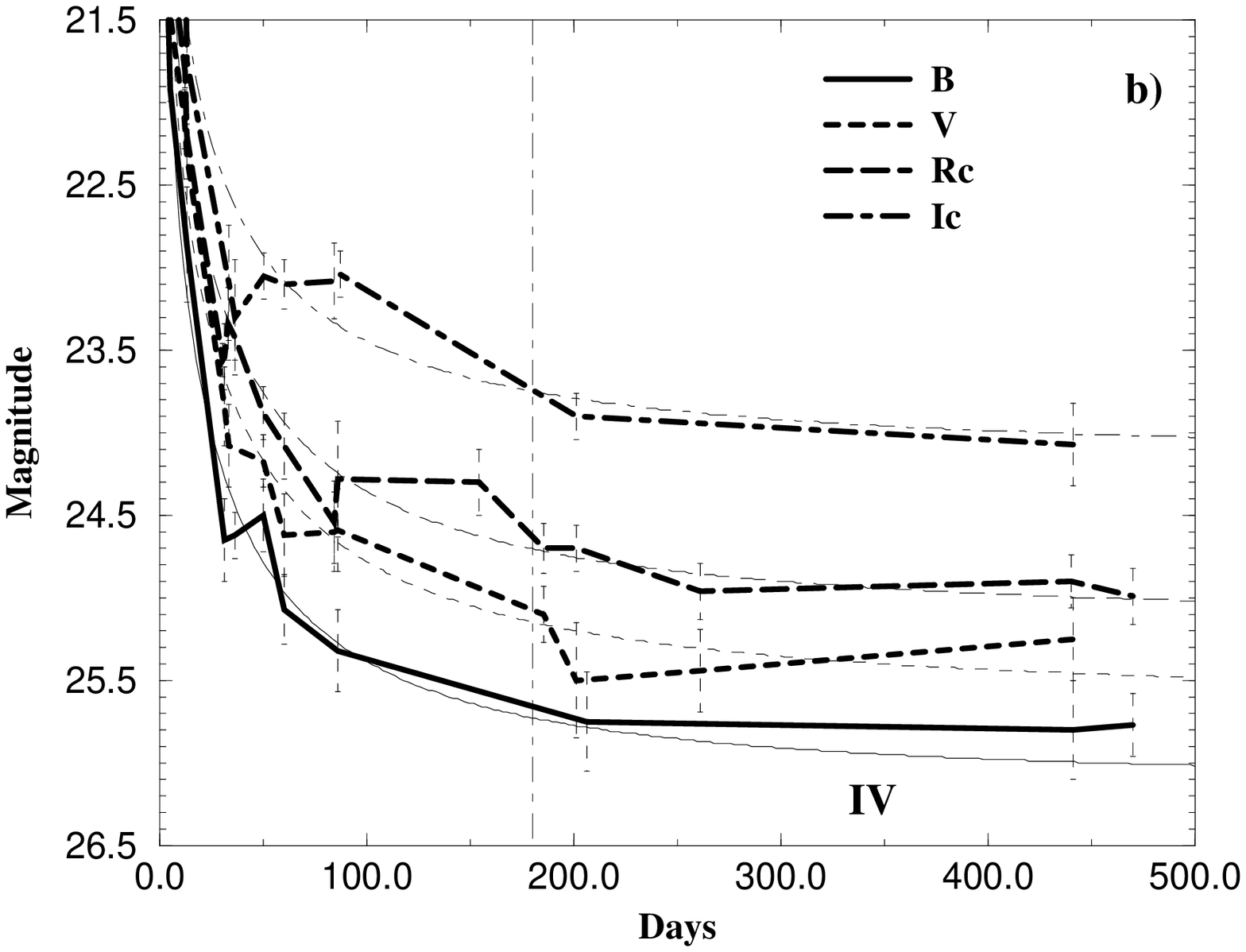}
}
\caption{
The behavior of $BVR_cI_c$
light curves of the OT + host galaxy of GRB 970508 up to about
470th day from the time of the GRB. Four independent
$BVR_cI_c$ power-law fits with different ${\alpha}$ (see text) are
exhibited by the thin lines.
The $BVR_cI_c$ light curves are shown on logarithmic ({\bf a}) and
linear ({\bf b}) scales.
}
\label{BVRIlight_curve}
\end{figure}

The OT of GRB 970508 is a second optical source related to a $\gamma$-ray burst
registered  by  the BeppoSAX satellite.
The optical variable object was first reported by H. Bond as a possible
optical counterpart of GRB 970508 (Bond, 1997) and was independently
found in our data ($R_c$ band, 1-m telescope) only about 0.5 day later.
Observations of the GRB 970508 optical remnant were continued
with the 6-m telescope in  the standard $BVR_cI_c$ bands
till Aug. 1998.
In Fig. \ref{BVRIlight_curve} a,b the behavior of the OT light curves
+ host galaxy of GRB 970508 up to 470th day is shown;
{\bf a)} for the logarithmic time and
{\bf b)} for the linear time for days after the GRB.
The $BVR_cI_c$ magnitudes with their errors (joined by the broken heavy lines)
are pictured. Four independent power-laws (the smooth thin lines)
determined by fitting the observed $BVR_cI_c$ light curves
with a two-component model, the sum of the OT GRB 970508,
fading according to a power-law
(but after the brightness maximum in Fig. \ref{BVRIlight_curve} {\bf a}),
and a constant brightness host galaxy ($ F = F_o\times t^{\alpha}+F_c$
with different ${\alpha}$) are shown (for details see Sokolov et al., 1999).

The data obtained with the SAO RAS 1-m and 6-m telescopes allow us
to divide the GRB OT brigthness changes at least
into 4 stages (Fig. \ref{BVRIlight_curve} a,b): \\
I ) the increase of brigthness on a time scale of about one day; \\
II ) (after the maximum)
the exponential brightness fall from our observations during about
4 days with a stable broad-band spectrum; \\
III ) further brightness fading
according to a rough power-law;
but note that from the 31-36th days after the GRB the object did not show
any {\it smooth}  further power-law fall at least in the $I_c$ filter. \\
IV ) after approximately 200 days any power-law decay ceased and the OT
contribution to the host galaxy flux
was already less than the observational errors. \\

At the first two stages --- the increase of brigthness and
the OT brighness maximum --- the behavior of the light-curves
has not been understood yet (see Panaitescu \& Kumar, 2001).
Nevertheless, at the OT brighness maximum and immediately after it,
all $BVRI$ fluxes were measured with the smallest errors, 0.03 -- 0.07 mag.
And any power law does not best fit for the brightness fading
after the maximum (May 10.77 UT). We see the same exponential law in
all $BVRI$ bands (Sokolov et al., 1998), or
an identical exponential brightness fading
was observed in 4 bands simultaneously.

After the 5th day the light curves in the $BVR_cI_c$ bands can be
described by power laws
(with different ${\alpha}$).
But starting with about the 31st day
after the GRB, the OT did not show further fall of brightness
in the $I_c$ filter. This phenomenon lasted not less than 51 days.
In that period the OT was stable around $I_c$ = 23. After 87 days
the deviation from the "average" power law
was already as high as $\simeq 1$ mag.
($I_c$  was  23.04$\pm$0.14 for the 87.085th day after the GRB).
The behavior we observed can also be explained by some intrinsic
features of GRB 970508 OT.
In this temporal and spectral behavior of the GRB OT there
can be a hint of a similar behavior of supernovae (see below).

Late-time observations of the GRB 970508 optical counterpart in
July-August 1998 show that the optical counterpart did not change
during the last half-year, and we can conclude that we observe a pure GRB
host galaxy with
$\ B = 25.77\pm 0.19, $ $\ V = 25.25\pm 0.22 $,
$\ {R_c} = 24.99\pm 0.17 $, $\ {I_c} = 24.07\pm 0.25 $,
already without the GRB optical remnant contribution.
Comparison of the $BVRI$ spectral energy distribution of the GRB host
with local starburst galaxies leads to the best fits for Scd starburst
H II galaxies (Sokolov et al., 1999; 2001).

The global best $\chi^2$-fits
(see Fig. \ref{BVRIlight_curve})
with a single power law, for most of the filters used,
are not acceptable (Sokolov et al. 1998).
Some intrinsic variability of the GRB optical remnant
fading is possible, which demands a more complex law presenting
the largest multiband homogeneous data set from one telescope.

\section{Supernovae and Gamma-ray beaming}\label{Beam}

The $I_c$ band light curve of GRB 970508 afterglow exhibits a rebrightening
phase, or a "shoulder" (Fig. \ref{BVRIlight_curve} a,b).
This spectral property
lasts for not less than 51 days.
Over across this period the color index $(V-I)$ increases
up to $\approx$1.6 mag ($I_c = 23.04 \pm 0.14$ for the OT).
The deviation from the "average" power law
with $<{\alpha}> = -1.25$ exceeds 1 magnitude.
Recent investigations
show that GRB optical afterglows have unusual temporal and
$VRI$ spectral properties: GRB 970228 (Galama et al., 2000),
GRB 980326 (Bloom et al., 1999) , GRB 990712 (Bjornsson et al., 2001),
and GRB 991208 (Castro-Tirado et al., 2001).
All these OTs have temporal and spectral characteristics
similar to those of GRB 970508, for which the $I_c$ band light curve is
consistent with the assumption that the SN emission overtakes the OT flux at
late times, i.e. nearly 31 days after the GRB. It should be noted that the
emission of all Type I SNe shows a strong UV deficit owing to absorption
lines below 3900\AA. Thus, for the redshift $z = 0.835$
a weak flux is expected from the suggested SN in the $R_c$ band
corresponding to the $U_{rest}$ (3652\AA) in the rest frame.

In contrast, the SN brightness must be
in the maximum at larger wavelengths (Fig.1),
i.e. in the $I_c$ band as observed for GRB 970508
(and for GRB 970228 with $z = 0.695$ (Galama et al., 2000) plus
time profile stretching for the light curve of the SN by a factor of ($1+z$).
One can see the effect from an "average" power-law fit
$ F = F_o\times t^{\alpha}+F_c$ with $<{\alpha}> = -1.25$.
The deviations ($\chi^2$/d.o.f) from the "average" power-law increase
toward the red part of the OT $BVR_cI_c$ spectrum:
$\chi^2$/d.o.f = 1.57 for the $B$-band, 2.41
for the $V$-band, 2.93 for the $R_c$-band, and 3.62 for the $I_c$-band
(see Sokolov et al., 1999 for details).

As $z = 0.835$, the $I_c$ band
in the observer frame corresponds to the $B_{rest}$ (4448\AA)
band in the rest frame,
the observed $I_c$ flux is thus used to derive the peak absolute magnitude
($M_B$) of the assumed SN. At the $I_c$ "shoulder" (Fig. 2)
the brightness of GRB 970508 OT corresponds to the maximum
luminosity of a peculiar SN with $M_B$ around -19.5
if for GRB 990708 host we have $M_{B_{rest}}$ around -18.5
(Sokolov et al., 1999).
We can consider type-Ib/c SNe as
a preliminary model of GRB/SN $\gamma$-ray bursters, but the peak magnitudes
of type-Ib/c (or core collapse) SNe are not constant: $M_B$ is
from -16 to -19.5.
Thus, the GRB host galaxies (for ground-based observations,
at least) must be dimmer than the peak magnitude of the SN (see Table 1 in
Sokolov et al., 2001).

SN 1997ef has been recognized as a peculiar SN Ic from its light curve
(Iwamoto et al., 2000).
It shows a very broad/flat peak and a slow tail.
(But the luminosity is lower than that of the well-known SN 1998bw.)
However, for the purposes
of this paper, the shape of SN Ic light curve is sufficient as an example
of type-Ib/c SN.
We do not know exactly the light curve (and the spectrum)
of the true SN accompanying GRBs; may be in the OT GRB 970508
case it was a peculiar SN by its luminosity and by the shape of
the light curve (Fig. 2).


\begin{figure*}
\centering
\includegraphics[width=0.75\textwidth,bb=30 30 720 522,clip]{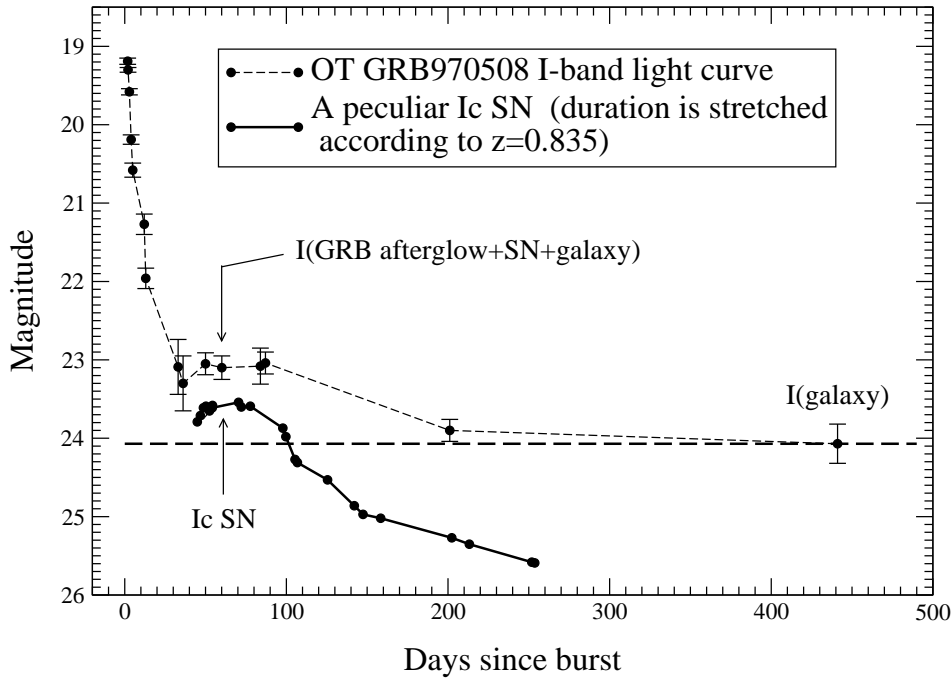}
\caption{
OT GRB~970508 $I$-band light curve plus a kind of a peculiar Ic SN
similar to SN 1997ef by the shape of the light curve
(but with the peak luminosity like SN 1998bw).
}
\label{OT_GRB970508}
\end{figure*}

So, we can expect GRBs to originate from massive stars which have lost their
hydrogen envelope before a SN-like explosion.
If GRBs arise from collapses of massive stars in starforming galaxies,
it is an unavoidable consequence that emission from an underlying SN should be
superimposed on the GRB afterglow.
Economy of hypotheses leads us
to suppose that all or at least the main part of long duration
{\bf GRBs are associated with SNe} (but not all SNe are associated with GRBs).

From the observations of 1997-2000 for known GRB host galaxies we have
(see Fig. 3 ):
$$m_{HostGal}= \mbox{from} \sim 22 \mbox{ to } \sim 28 \mbox{ st.mag.}$$
Such faint (or fainter) galaxies are indeed expected from studies
of the properties of the GRB host galaxies.
It leads to the fact that
the search for direct GRB-SN associations in galaxies close to us
is a challenge, because the majority of the SNe related to GRBs
will be faint (22-26 mag) and also in very distant galaxies with $z\ge 0.4-4.5$.

\begin{figure}
  \resizebox{\hsize}{!}{\includegraphics[clip]{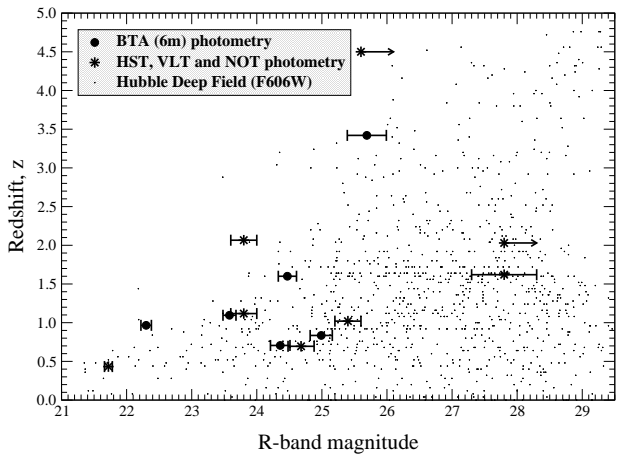}}
\vspace{-0.25cm}

\caption{
Observed $R$-band magnitude vs. redshift of the GRB host galaxies.
The $R$-band magnitudes  are  corrected  for  the  Galactic extinction.
The BTA $R$-band magnitudes (from Sokolov et al., 2001)
are marked with circles, while asterisks refer to the results of other authors.
Also the HDF F606W magnitude vs. photometrical redshift
distribution is plotted. Catalog of the F606W magnitudes and photometrical
redshifts was used from Fern\'andez-Soto et al., 1999
}
\vspace{-0.5cm}
\label{R-z_diagram}
\end{figure}

Having established that the GRB hosts are faint,
for a number of galaxies brighter than 26 mag
in one square degree of the sky we have
$N_{gal < 26 mag} \approx 2\cdot 10^{5}\cdot gal\cdot deg^{-2}$
(Casertano et al., 2000).
Taking into account the observations with BATSE
$ n_{GRB} \approx 0.01\cdot deg^{-2} yr^{-1} $,
we conclude that the rate of GRB events is
$$N_{GRB} \approx 5\cdot 10^{-8}\cdot yr^{-1} galaxy^{-1}.$$
For the SNe rate in the local starburst galaxies we have
$$ N_{SN} \approx 0.02\cdot yr^{-1} galaxy^{-1}$$
for  all types of SNe combined.
It should be kept in mind that GRBs are not observed in all galaxies and
that not all types of SNe can be related to GRBs, but only type-Ib/c
(Capellaro et al., 1999).
So we have $ N_{SN} \approx 0.001\cdot yr^{-1} galaxy^{-1}$.
But if one looks at the matter the other way round, we must remember:
the massive star-forming in very distant (and dusty) GRB galaxies
with $z\sim > 1$
is more vigorous than in the local starburst galaxies or in galaxies
like the Milky Way (Blain and Natarajan, 2000),
giving rise SN bursts of 10 times and more.

Thus, with due regard for these reservations, an estimate for
the number of GRBs which could be related to SNe
is of order of
$$N_{GRB}/N_{SN}\approx 5\cdot (10^{-5} - 10^{-6}).$$
%
So, if some GRB -- SN relation really exists, then we have two possibilities:

{\bf 1)} either we deal with a very very rare type of SNe (or "hypernova"?) --
GRBs related only to the $10^{-5}-10^{-6}$-th part of all observed SNe
in distant galaxies (up to 28th mag), or

{\bf 2)} we have a very strong {\bf $\gamma$-ray beaming} with a solid
angle up to
$$\Omega_{beam} \sim (10^{-5} - 10^{-6})\cdot 4\pi,$$
%
if GRBs are associated with an {\bf asymmetric} SN explosion,
whereas the $\gamma$-ray beam (and narrow collimated jet)
is directed towards the observer on the Earth.
And if GRBs are so highly collimated, radiating only into a small fraction
of the sky, the enegry of each event may be much reduced, by several
($\sim 5 - 6$) orders of magnitude:
$$E_{beam} = E_{\gamma} \Omega_{beam}/4\pi \sim 10^{45} - 10^{47} ergs $$
for the isotropic equivalent of the total GRB energy
$E_{\gamma} \sim 10^{51} - 10^{52}$ ergs.

Thus, if all or the main part of long GRBs are associated with SNe,
we need more examples to test the GRB-SN link. But from the Earth we can see
only peculiar SNe (by their luminosities and by light curves) connected with
GRBs. The SNe must be brighter than their hosts. The luminosity of these dusty
galaxies $M_B$ is basically between -18.6 and -21.3
(Sokolov et al., 2001).

\section{Discussion: still, what are the progenitors of Gamma-Ray Bursts?}
\label{Disc}

\subsection{X-ray spectra and close environment of GRB source}\label{X-ray}

The BeppoSAX, ASCA and Chandra observations in X-rays confirm rather
a link with the death of massive stars; the satellites found a spectral
signature of iron from different GRBs.
To date, four GRBs have shown evidence
for an iron emission line during the X-ray
afterglow, observed 8-40 hours after the burst event (GRB 970508, Piro et al.,
1999; GRB 970828, Yoshida et al., 2001; GRB 991216, Piro et al., 2000;
GRB 000214, Antonelli et al., 2000).
SNe or SN-progenitors can be the most plausible source of such an iron.
In any case, the new data are very important, because all spectral
emission lines observed so far are produced by the GRB host galaxies
rather than at the burst site.

   Chandra's observation of the brightest GRB 991216 was started
37 hours after the BATSE/GRBM trigger (Piro et al., 2000). The spectrum of
the X-ray afterglow shows the emission line at energy $E = 3.49\pm 0.06$ keV.
The emission feature is associated with the rest energy of 6.97 keV of the
emission of H-like ions of iron (FeXXVI) because at higher energies the
Chandra/ACIS-S spectrum shows evidence of a recombination edge in
the emission at $E = 4.4\pm 0.5$ keV.
The identification of this feature with the iron recombination edge
(with the rest energy of 9.28 keV)
gives $z = 1.11\pm 0.11$ consistent with $z = 1.02$
found in optical spectroscopy of the OT GRB 991216 (Vreeswijk et al., 2000).

An iron recombination edge at 9.28 keV is, indeed, expected when the iron
emission is driven by photoionization and the medium is heavily ionized by the
radiation produced by X-rays from the GRB or its afterglow.
If the medium lies
in the line of sight, the edge is expected to be seen in absorption at early
times, before the photoionization of the medium. (Evidence of such a
feature has been found in another GRB, GRB 990705.)
At later times, when the medium
becomes heavily ionized and recombination takes place, the edge is seen in
emission. In this condition X-ray lines are produced almost exclusively
through recombination of electrons on H-like iron.
From the Chandra observation of GRB 991216,
the distance of the medium from the GRB source or the size of the region of
the emission feature formation around the GRB site is
$R > 2\cdot 10^{15}$ cm ($R = c\cdot t$). The latter condition is derived
from the presence of the line 1.5 days ($> 37$ hours) after the GRB,
which is shorter than the observed one by a factor of $(1+z)^{-1}$.

Different scenarios have been proposed to explain the lines in emission
(see Lazzatti et al., 2001; Vietri et al., 2001; Rees and Meszaros, 2000;
Weth et al., 2000). The observations by BeppoSAX, ASCA and
Chandra seem to indicate large amount of iron close to the GRB site.
One of the explanations emerging (Vietri et al., 2001)
from all pieces of evidence is as follows:
a massive progenitor --- like a "hypernova" (Paczynski, 1998)
or a "collapsar" (Woosley, 1993) --- ejects, shortrly
(several months) before the GRB, a substantial fraction of its mass.
But there is room for a lot of models so far
and (for all that)
other more simple scenarios
invoking a simultaneous explosion of a GRB source and a progenitor star
are possible --- see below.

Unlike of GRB 991216 and others which showed evidence for the iron emission
lines during the X-ray afterglow, GRB 990705 showed a
prominent {\bf absorption} feature at 3.8 keV and an equivalent hydrogen column
density that disappeared 13 s after the burst onset (Amati et al., 2000).
The absorption feature was interpreted as due to an edge produced by
the neutral iron (i.e. at 7.1 keV),
redshifted to 3.8 keV (corresponding to $z = 0.86$).
And what is the most important:
the absorption feature seen during the first 13 seconds
from the BeppoSAX/GRBM trigger, became undetectable afterwords.
The straightforward (or a "simplest") interpretation of the absorption feature
reported by Amati et al. (2000) is that neutral iron is present along the
line of sight to the burst. This iron is photoionized by the prompt GRB
X-ray photons untill all electrons are stripped from the iron ions,
causing the disappearance of the edge. But in this case the size
of the region of the absorption feature formation around the GRB site is
($l = c\cdot t$) $l \approx 4\cdot 10^{11} $cm $\approx 6 R_\odot$.
It is quite immediate environment of the GRB site!

Emission and absorption features in the X-ray spectra of GRBs and their
afterglows provide a fundamental new tool to study their close environment
(or the circumburst) and thus their possible progenitors.
Theoretical computations (Ramirez-Ruiz et al., 2001) show that only
a {\bf dense}, massive medium close to the GRB site --- such as that
expected in the case of a massive progenitor --- could produce an iron
emission line detectable with the current X-ray instrumentation.
Ramirez-Ruiz et al. (2001) argued that high-density shells could be produced
from a low-density and low-velocity wind if the massive
progenitor or a massive star undergoes a Wolf-Rayet (WR) phase, which is
characterized by a strong stellar wind that causes the star to lose enough of
its outer layers for the surface hydrogen abundance to become minimal.

The radius of a WR star is sufficiently small before the internal engine
ceases to operate. Because of the intrinsic variation of mass-loss rates
in WR evolution,
we have shells of varying gas density. The effects of the WR ejecta interacting
with the interstellar medium (ISM)
can be observed in the wind bubbles around some
of these objects. The deceleration of a pre-SN wind by the pressure of the
surrounding medium or the interaction of fast and slow winds could also
create circumstellar shells.
Moreover, massive stars tend to lie in associations, so there is the
combined effect of the winds of many stars.
Thus,
the state of the ambient medium of a WR star {\bf at the end of its life} is
determined primarily by the time evolution (until core collapse) of the
mass-loss rate. Here, by the paper of Ramirez-Ruiz et al. (2001),
we emphasize some interesting contributions to the structure of the
ambient medium that arise when the ejected winds interact with the ISM
during the WR lifetime (about $10^{5}$ yr).

From the inside (near WR site) and to the outside: the external medium particle
density of the wind ejected by the progenitor is best modeled by a power-law
$n = A r^{-2}$. A spherical shell with large column density
arises when the ejected
material accumulates at the radius $r_{dec}$ at which the free expansion
phase of the ejected wind terminates. So, we have set the density of the
ambient material $n_0$  before the WR phase to be 1 cm$^{-3}$.
But if a neighbouring star is present, the density ($n_0$) of the surrounding
medium can be as large as $10^{6}$ cm$^{-3}$, causing the ejected wind to slow
rapidly at a much smaller radius, $r_{dec}$.
Figs 2b and 4b from the paper of Ramirez-Ruiz et al. (2001)
show the effect of different initial masses on the structure of the ambient
medium. The density can be fited by the power law $n(r) = Ar^{-2}$ for
$r < r_{dec}$. The value of the constant $A=3.0\cdot10^{35}A_{*}$
as a function of $M_{0}$ is shown in panels $a$.
The Figs. 2 and 4 show (panels $b$)
also the radial location ($r_{dec}$)
of the highest density shell as a function of initial main-sequence
mass $M_0$ for different values of $n_0$ and for different metallicity
(Fig. 4 in the Ramirez-Ruiz's et al. paper).
The free expansion of the ejected wind is terminated when the swept-up mass
becomes comparable to the mass of the wind. The ejected mass then accumulates
at the radius $r_{dec}$. This overdense region lies closer to the progenitor for
higher density $n_0$ of the surrounding medium (or the density of ambient
material before the WR phase) and for WR stars originating from low initial
mass stars $M_0$.
Panels $c$ in Figs. 2 and 4 (Ramirez-Ruiz et al., 2001):
all the evolved WR stars have lost their hydrogen envelope
(Bisnovatyi-Kogan and Nadyozhin, 1972)
and are left with a bare helium core prior to core collapse (or prior to GRB?).

An important result of these calculations for us is the fact that
for the initial star masses of $M_{0} \approx 30M_\odot $ the region,
in which almost all mass ($\sim 10-20M_\odot $) released during the WR phase
is concentrated, can really be of a typical dimension of $\sim 10^{15}$cm,
which is also the result of BeppoSAX, ASCA, and Chandra's
X-ray observations of the GRBs (GRB 970508, GRB 970828, GRB 991216,
GRB 000214).
If there exists the overdense region ("bump") or not
depends on the density of the ambient material $n_0$ before the WR phase.
(Note -- it can be important for the OT GRB 970508, --
many WR stars appear to be single; however, the fraction of visible close
WR + OB binaries seems to be around 35 per cent.)
The internal (or the inside) medium particle density
near the GRB progenitor follows the power-law $n = A r^{-2}$
where $A = A_{*}\cdot 3\cdot 10^{35}$cm$^{-1}$ and $0.1 < A_{*} < 10$
for WR stars at the end of their lives.
Besides, there can exist a region of dimension of
$2r_{dec}\sim 10^{15}$cm or a more or less sharp boundary in which
the released mass stopped by the interaction between the WR winds and
the ISM.
Apparently, such a boundary around the WR star or the GRB progenitors
should arise in typical star-forming regions with densities of
at least and perhaps significantly exceeding 10 - 100 cm$^{-3}$,
which is characteristic of large molecular clouds complexes if
GRBs are associated with massive stars indeed.

\subsection{Optical and X-ray light curves of the OT of GRB 970508}
\label{Curves}

Panaitescu, \& Kumar (2001) analyzed
the afterglows of four {\bf well-studied} GRBs
(GRB 980703, GRB 990123, GRB 990510, GRB 991216),
assuming that the emission is due to the interaction
of a collimated relativistic shock with the ambient medium.
They have shown from their particular afterglow shock model
that the ambient density of some GRBs can be as low as
$\sim 10^{-1} - 10^{-4}$ cm$^{-3}$ by contrast with densities, 10 - 100
cm$^{-3}$ (and perhaps even more), that should be typical for the star-forming
regions. But we have to note that the ascribed {\bf paradoxical}
densities are very model dependent.
We have now the straightforward data from X-ray spectra:
{\it e.g.} GRB 991216 just from their GRB list.
(The search for spectral features of one of the
primary objectives of Chandra and BeppoSAX's GRB observation programs.)
In addition, it should be remarked that Panaitescu \& Kumar (2001)
in the analysis of multi-wavelength data of four GRBs leave out
the {\bf best-studied} OT of GRB 970508,
whose optical light-curve cannot be entirely explained within the
framework of their GRB afterglow shock model,
as the OT exibited a re-brightening after 1 day,
indicating a possible delayed energy injection or a strong
fluctuation of the external density or
the fluctuation of just the same ambient density (or $n_{0}$ ?).

Piro et al. (1999) have searched the X-ray spectrum of GRB 970508's afterglow
for an Fe line observed by BeppoSAX. They found such a line but with a limited
statistical significance (99.3 \%) in the early part (first 16 $h$) of the
afterglow. The line decreases in the later part of the observations ($\sim 1$
day after the GRB) by at least a factor 2, enough to make it undetectable with
current apparaturs. Simultaneously with the line disappearance, the X-ray
flux both rises (see Fig. 5 $(b)$)
and hardens consistent with the appearance
of a new shock (?). Then, at the end of the outburst, the spectrum steepens.
It is most likely that the line of GRB 970508 is produced by
fluorescence and recombination of Fe atoms ionized by the intense X-ray flux
of the GRB.
Judging by the X-ray and $\gamma$-ray peak fluxes (Frontera et al., 2000)
the prompt X-ray radiation simultaneously detected with the BeppoSAX/GRBM
(40-700 keV) and the BeppoSAX/WFC (1.5-26 keV)
can produce up to $10\%$ of the total energy of GRBs.
%
So, at the early phases of the GRB the radiation
field is so high ($F_{X}\sim 10^{17} - 10^{18}$ erg cm$^{-2}$ s$^{-1}$
for $r = 3\cdot 10^{15}$cm) that iron atoms are
completely (or nearly) stripped of their electrons.
%
From the line variability for GRB 970508
one can deduce that this ambient medium should be located at
the distance of
$c\cdot t \approx 3\cdot 10^{15}$cm from the GRB source.

That is all at the moment what can be said about the interpretation
of the X-ray observations, independetly of one or another model
adhered by the authors themselves
(Piro et al., 1999; see also Vietri et al., 2001).
In particular, the estimate of the mass of Fe can be model-dependent.
Note that the GRB beaming would introduce an additional geometrical factor.
This would reduce the complete number of the X-ray's photons and hence
reduce the mass of Fe required to produce the emissiom line.
We can try to account for this circumstance after comparison of
X-ray and optical data for OT GRB 970508.
But we can agree at once with the authors
(Piro et al., 1999; Vietri et al., 2001) in one thing,
independently of the model in which the mass of Fe is calculated:
in contrast to the fairly clean environment expected in the merging of two
neutron stars, the observed line properties would imply that the site of the
GRB is embeded in a large mass of material, consistent with some
pre-GRB-explosion ejecta arising from a massive star.

Besides, if the scenario
{\bf"a massive star ---$>$ WR ---$>$ pre-SN = pre-GRB"}
with the formation of a shell around WR is true, then it could be supposed
that the reason of the arising of relativistic jet is the powerful light
pressure of the prompt X-ray radiation of a GRB source.
For the force of the light pressure acting on the
environment around the GRB source (WR-star ?)
   we have $L_{GRB}/(4\pi r^{2}) \cdot (\sigma_{T}/c)$,
where $L_{GRB}$ is {\bf an isotropic equivalent
  ($\sim 10^{50-51}$erg$\cdot$s$^{-1}$)} of GRB luminosity,
$r$ is the distance from the center (near the GRB site),
   $\sigma_{T} = 0.66 \cdot 10^{-24}$cm$^2$ is the Thompson cross section,
$c$ -- velocity of light.
Even withouth detailed calculation it is clear that such a force
can exceed the Eddington limit
  ($\sim 10^{38}$erg$\cdot$s$^{-1}$ for 1 $M_\odot$) by 12-13 orders.
If we assume the GRB radiation (the prompt X-rays + $\gamma$-rays) to be
collimated or we have very strong beaming with the solid angle
   $\Omega_{beam} \sim (10^{-5} - 10^{-6})\cdot 4\pi$,
then the formation of an ultra-relativistic jet becomes inevitable.
Now everything depends on the ionization, density, and temperature
of matter in the immediate vicinity of the GRB source ---
an asymmetric collapsing nucleus of a massive star.
But then the behavior of X-ray and optical lightcurves for the
afterglow of GRB 970508 (see Figs. 4 and 5), inexplicable
within the framework of Panaitescu and Kumar's (2001) afterglow model,
becomes quite natural.

From Fig. 4 we see some deviations from an average
(or from a "standard") power law  in all 4 bands.
The broad-band optical spectrum of the OT did not change at all
in that period of the exponential brightness fall.
The exponential $B$, $V$, $R$, $I$ behavior of the OT immediately after
the maximum (Fig. 5 $a$) correlates with GRB~970508 X-ray
transient light curve (Fig. 5 $b$) after the obvious minimum
(Piro et al., 1998),
and the subsequent X-ray flux fading deviates from the power law.
The same as the OT decay does not follow any power-law (Fig. 4).
This is in favor of a common origin of the X-ray and optical events.
More than 16\%:  6\% (in optical) + 10\% (or $\sim 10^{50}$ergs in X-rays only)
of energy of the GRB itself was released in that period.
There was some activity which had rather high energy,
comparable to that of the GRB burst.
And it was on a time scale ($\sim 10^{5}$ s) 10.000 times longer than
the duration of the GRB itself.

This enigmatic re-brightening of the GRB 970508 afterglow
could be a consequence of collision
of the jet shock fronts with the high density spherical shell
or a sharp boundary (or the overdense region)
with dimension of $2r_{dec}\sim 10^{15}$ cm
in which the released mass stopped during the WR phase of the GRB progenitor.
That dense outer shell with a dimension of $\sim 10^{15}$ cm
catches up the essential X-ray energy and we have a new
X-ray source of 0.8 $day$ after the GRB (Fig. 5 $b$).
But we do not observe the jet itself. We see (especially in this maximum
brightness of GRB 970508 afterglow) the manifestation of both these shocks
and (as their effect) a powerful X-ray source arising as the jet moves
through the density inhomogenity
($\sim 10^{10}$ cm$^{-3}$) around pre-SN (= WR).
  Apparently, this is just the exponential decay lag of this shell
(a spherical layer of a thickness of $\sim 10^{15}$ cm too)
against the background of synchrotron radiation of the shock wave in it
that we observe during several days after the maximum (Fig. 4 and 5)
for the GRB 970508 afterglow.
After this a new X-ray source has flashed out (4-5 days after GRB),
only synchrotron radiation of the shock wave remains at passage
of the jet through the homogeneous ISM with a density of $n_{0}$,
which existed around GRB progenitor or around the WR star.
These are approximately the same mechanisms of radiation that
are used by Panaitescu \& Kumar (2001).
But perhaps the strong
deceleration due to the interaction of the relativistic
shock with the ambient medium does not arise as in their model
because the compact relativistic jet (or "bullet") is decelerating
but not the shock.
This "bullet" moves at a relativistic speed and
the shock (arising as it moves through ISM) only heats this medium.

It is not always that such an overdensity can arise around the GRB progenitors
unlike the case of GRB 970508 progenitor. It depends on the value of
$n_{0}$ and other factors (see Ramirez-Ruiz et al., 2001).
But apparently, a sphere of a size of $\sim 10^{15}$ cm
 around WR stars often forms
with a more or less sharp jump in density at its boundary.
In connection with that I make bold to say here some "seditious" thought:
the known breaks or steepening
of some GRB afterglow light-curves (have been
interpreted as a collimation effect in GRB-models with
fireball) can be explained in this scenario
(with the acceleration
of the relativistic jet by X-rays + $\gamma$-rays)
by the same interaction between this jet (or "bullet")
and the inhomogeneous medium around WR.
As the jet goes out of this sphere the medium density around
the GRB source changes and the sphere flashes on.
%
The breaks will appear at light curves that are related only to
inhomogeneities of $n(r)$ near the GRB progenitor.
(As to the breaks in the light curves see also
Pian et al. (2000), Dai and Cheng (2001), and references therein.)

And where does the acceleration of the jet up to a relativistic velocity occur?
Now we will try to answer this question.
The photon flux producing the radiation pressure accelerating the matter
at a distance of $r$ from the center (near the GRB site) is
$L_{GRB}/(4\pi r^{2})$.
It is inside this region and in the immediate vicinity of a GRB source
the photon flux turns out to be huge and at least not less
than Eddington's flux.
(From definition:
if $L_{Edd}$ is the Eddington limit $\sim 10^{38}$erg$\cdot$s$^{-1}$
for 1 $M_\odot$, and $R_{*}$ is the size of a compact object of
$\sim 10^{6}$ cm
then $L_{Edd}/(4\pi R_{*}^{2})$ is a flux stopping the accretion onto
a compact source --- the falling of matter on the source at
a parabolic velocity.)
From the condition that the photon flux
$L_{GRB}/(4\pi r^{2})$
at a distance of $r$ is equal to the $L_{Edd}/(4\pi R_{*}^{2})$
and taking into account that the luminosity of the GRB radiation is
$L_{GRB}\sim 10^{50-51}$erg$\cdot$s$^{-1}$ (an isotropic equivalent),
it is possible to obtain an estimate of the size ( $l$ ) of the region in which
the light pressure accelerates the matter around the GRB source to
relativistic and ultrarelativistic (for $r < l$ ) velocities:
$l\sim 10^{12}$ cm $\approx 14R_\odot$.
So, it can be near the region of a size of
$\approx 6 R_\odot$, which follows from the observation
of the variable absorption feature in the prompt X-ray emission of GRB 990705
(Amati et al., 2000).

\begin{figure}
\includegraphics[width=\hsize,bb=45 380 570 765,clip]{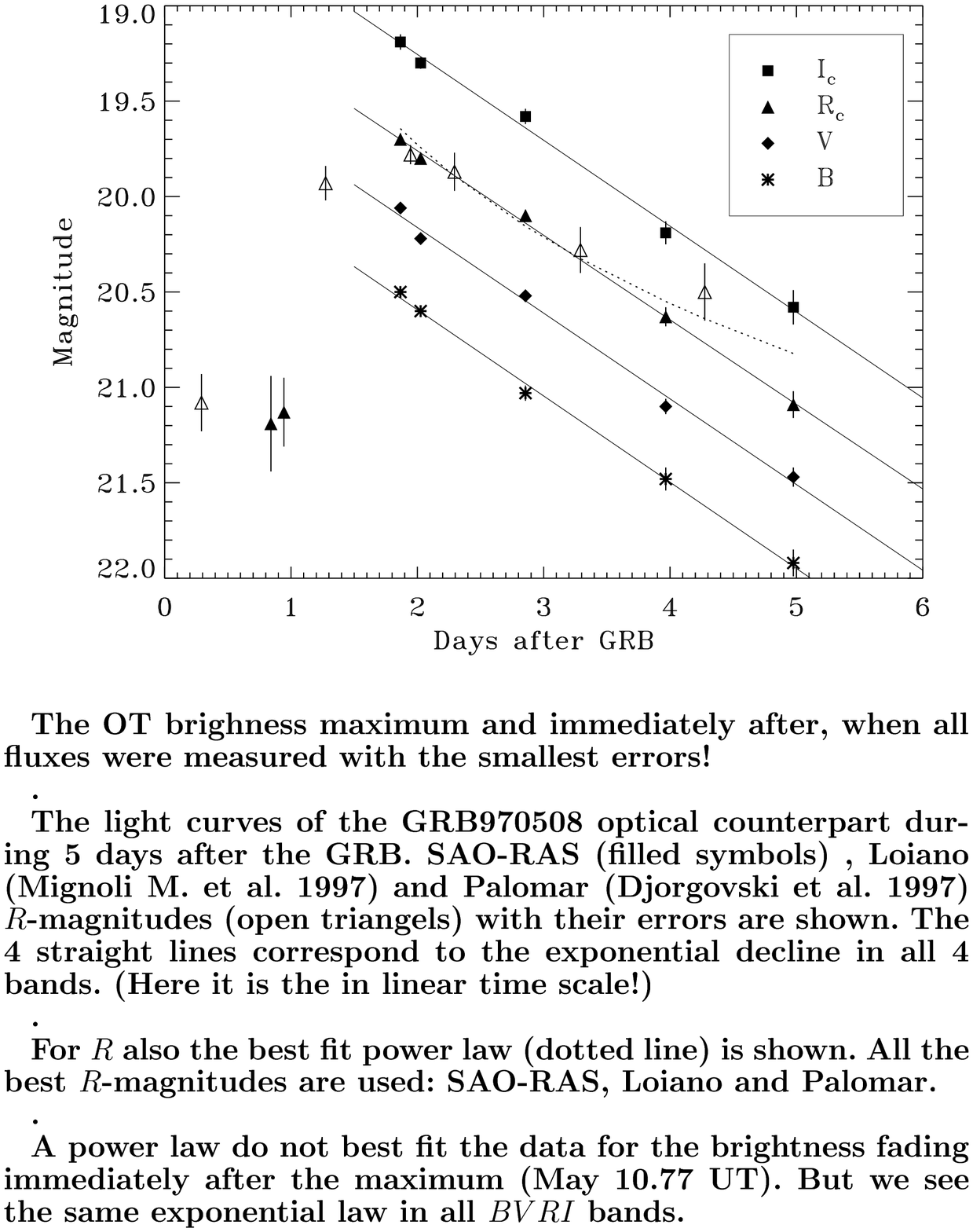}
\caption{
The light curves of GRB 970508 optical counterpart during
5 days after the burst. SAO RAS (filled symbols)
and Palomar
(open triangles) magnitudes with their errors are shown.
The lines correspond to the exponential decline of brightness
reported in (Sokolov et al., 1998).
For $R_c$ also the best fit power law is shown
(dotted line).
}
\end{figure}

\begin{figure}
\includegraphics[width=\hsize,bb=40 155 550 772,clip]{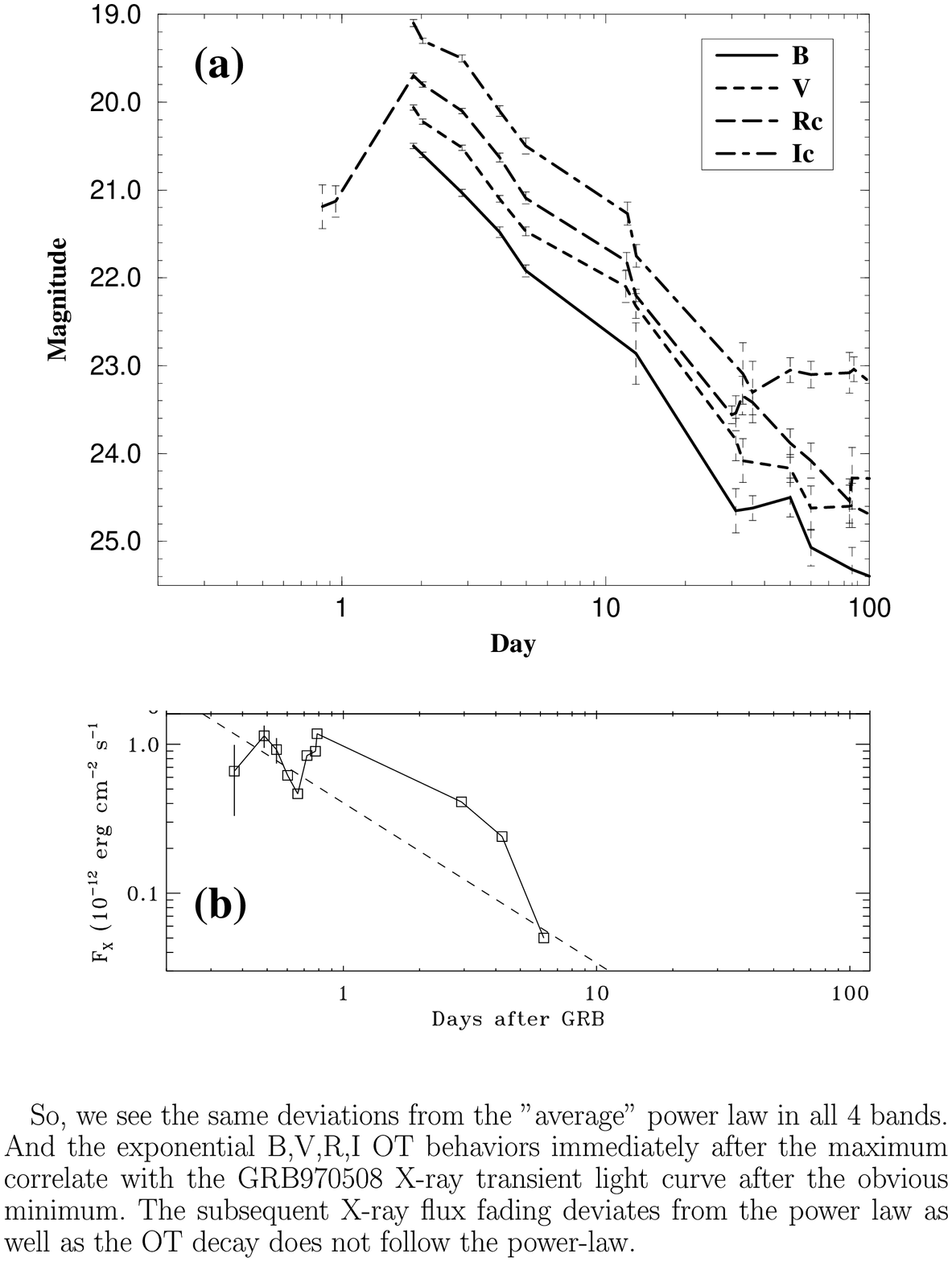}
\caption{
GRB 970508 light curves: {\bf (a)} $BVR_cI_c$ SAO RAS data;
{\bf (b)} 2--10 keV BeppoSAX MECS X-ray data
(Piro et al., 1998), the power law with the slope of $-1.1\pm0.1$
is shown (dashed line).
}
\end{figure}

\section{Summary}\label{Sum}

There is circumstantial evidence that GRBs are associated with the collaps
of massive stars.
The observations in optical (GRB host galaxies, OT light curves) and
GRB X-rays  (light curves and spectra)
suggest that this is so.
Nevertheless,
does every long GRB begin with the core-collapse supernova explosion?
It is still early to draw firm conslusions --- new
observations are needed --- but in any case a link between GRBs and SNe seems
to be convincingly confirmed.
Weak emission line features 8-40 hours  after the GRB may also result
from the supernova GRB scenario. In all these cases significant X-ray absorption
features, in particular during the prompt GRB phase, are expected. No
significant X-ray spectral features should result from
compact-object binary mergers.

Here an attempt is undertaken to understand
the light curves (in X-ray and in optical)
of the OT of GRB 970508
which have never been explained properly so far.
The X-ray spectra and the light curves (like OT GRB 970508) some of
GRB afterglows point to the existence of
{\bf the dense medium surrounding GRB sources }, which,
combined with the data on GRB host galaxies and the manifestations of SNs
on the light curves, agrees well with the scenarios where
the GRB progenitor (or pre-SN) is a WR star.

If a GRB progenitor always coincides with a pre-SN progenitor
(or GRB progenitors are actually WR stars), then the GRB radiation
is always very collimated. Then the formation of relativistic jets becomes
an inevitable consequence of huge radiational pressure owing to
the huge flux during the GRB.
We have not known so far how to produce such a narrow bunch
of hard X-rays and $\gamma$-rays
or what the mechanism of GRBs is. (The same as fireball GRB model makers
do not know how to produce it --- the fireball.)
But in that scenario,
with the {\bf straightforward} link of the GRBs and core-collapse SNe
and (as a result) with the strong $\gamma$-ray beaming,
the energetics of the GRB sources is substantially lower
and the energy release is not higher than the bolometric luminosity of SNe.
So, we can reduce the GRB problem to the old one --- the problem
of the SNe explosion mechanism. But this is a relief too ---
the number of hypotheses on the GRB sources does not increase at least.

\begin{acknowledgements}
I thank S.N.Fabrika and Yu.V.Baryshev for fruitful discussions
and for useful comments as to this manuscript.
The  work  was  supported by the ``Astronomy" Foundation (grant 97/1.2.6.4)
and RFBR N01-02-17106a
\end{acknowledgements}

%

\end{document}